# Cylindrical-water-resonator-based ultra-broadband microwave absorber

**JIAN REN,**[1, 3] **JIA YUAN YIN,**[2, 3*]

[1] *The State Key Laboratory of Millimeter Waves and Department of Electronic Engineering, City University of Hong Kong, Kowloon, 000000, Hong Kong SAR, China*

[2] *School of Physics and Optoelectronic Engineering, Xidian University, Xi'an, 710071, Shannxi, China*

[3] *These authors contributed equally to this work*

*email: jyyin@xidian.edu.cn

**Abstract:** A cylindrical-water-resonator-based absorber with ultra-broad operating band at microwave band is demonstrated theoretically and experimentally for the first time. By utilizing dielectric resonator (DR) mode and spoof surface plasmon polariton (SPP) mode of the cylindrical water resonators, the proposed absorber owns an absorptivity higher than 90% over almost the whole ultra-broad operating band from 5.58 GHZ to 24.21 GHz, with a relative bandwidth as high as 125%. The angular tolerance and thermal stability of the proposed absorber are simulated and the results indicate the good performance of the absorber under wide incident angles and weakly dependent on water temperature. Low cost, ultra-broad operating band, good wide-angle characteristic and thermal stability make the absorber promising in the application of antenna measurement, steady technology and energy harvesting.

## 1. Introduction

For a long time, electromagnetic (EM) absorbers have been attracted much attentions as its wide use and potential applications in areas of electromagnetic compatibility [1], sensors [2, 3], bolometers [4, 5], solar energy-harvesting [6, 7], thermal emitters [8-10], stealth technology [11-13], emerging passive cooling technologies [14-16] and so on. In this context, numerous methods have been investigated to obtain flexible control of different features of EM absorbers, like working frequency, absorptivity and polarization response. Among these features, operating bandwidth is one of the most concerned issues as the absorber with wide bandwidth is highly desired in different applications. To obtain wider operating bandwidth, multilayer structures [11] and lossy materials with tapered shape [17] are commonly used, which may result in bulk volume and high cost. In 2008, perfect metamaterial absorber (PMA) was first proposed by Landy and his colleagues [18], providing the possibility to design absorbers with extremely thin thickness. After that, many attempts have been focused on metamaterials/metasurface-based absorbers, from microwave band to optical band. Limited by resonating attribute, the operating bandwidth of PMA is usually narrow. To solve this issue, PMAs with dual- [19, 20] and multi- [21, 22] operating band have been investigated during past decade and the combination of these multi operating bands helps to obtain broadband feature [23, 24]. What's more, it has been proved that the absorption in PMA is mainly caused by dielectric loss [25, 26], implying that the use of materials with high dielectric loss has big potential in the design of near-unit absorption with wide operating band [27, 28].

Water, as one of the most accessible materials in nature, has been widely investigated in EM areas as its advantages of low cost, environmental friendly, and easy available. It has been widely used in the design of different microwave antennas, such as monopole antenna [29], reconfigurable antenna [30] and dielectric patch antennas [31, 32]. Recently, water has also become a hot spot in the design of all-dielectric metamaterials [33-35], which can support electric or/and magnetic resonance through properly tailoring the shape or orientation of dielectric resonators. And because of the fluidity of water, tunable all-dielectric metamaterials are easy to be achieved [36, 37].

Different from other microwave dielectrics, the imaginary part of water's permittivity, $\varepsilon''$, is quite large in microwave band, making it shows high dielectric loss and suitable for design of EM absorber with high absorptivity. At the same time, the real part of the permittivity, $\varepsilon'$, is largely depended on frequency, making it possible to obtain water-based microwave components with ultra-broad operating bandwidth. However, one issue that will limit the directly use of water as microwave absorber is the impedance mismatch at the interface of water and air, which will prevent EM waves propagate into the water, leading to low absorption. In this context, a plenty of previous works have been carried out to eliminate such mismatch. Take advantages of high dielectric loss of water, different water-based wideband EM absorbers have been proposed during past years. In 2015, Yoo *et al.* proposed a metamaterial absorber using periodic water droplets for the first time, covering 8-18 GHz [38]. Water-based metasurfaces with gridding shape were also designed in Refs. [39] and [40] for multiband coherent effect absorption and ultra-wideband absorption, respectively. Pang *et al.* put forward a thermally tunable broadband metamaterials absorber based on water substrate [41]. In [42], water resonator with sphere cap shape is presented for wide microwave absorption and dynamically tunable absorption by changing the height of the water resonator. In addition, water resonator in rectangular shape with a relative bandwidth of 28.8% has also been investigated [43]. Most recently, a water-based metamaterial absorber is designed by introducing a cylindrical hole in the water plate [44]. This kind of absorber has an ultra-broadband operating frequency from 12-29.6 GHz. Besides, the water cube is also studied for broadband microwave absorber [45].

In this paper, a cylindrical-water-resonator-based metasurface for microwave absorption is proposed. Both the dielectric resonator (DR) mode and the spoof surface plasmon polariton (SPP) mode of the cylindrical water resonator are excited, resulting in near unit absorption over an ultra-broad bandwidth. Different from the previous designed absorber based on spherical water resonator or water plate, here, the water resonator with cylindrical shape is investigated for the first time. Taking advantage of the frequency-dependent permittivity of the water in the band of interest, the water resonator can resonate in its DR mode over a wide band at lower frequency with two absorption peaks. While at higher frequency, spoof SPP mode of the water resonator can be excited over a wide band and another two absorption peaks appears. Through the optimization of the dimensions of water resonator, an ultra-broad absorption band can be achieved covering from 5.58 GHz to 24.21 GHz with absorptivity higher than

90%. The relative bandwidth reaches as high as 125%. Numerical simulation and experiment are carried out to verify the working principle of the designed ultra-broadband absorber. Good agreement between the simulated and measured results implies the validity of the proposed design. The angular tolerance of the absorber is also discussed, showing high absorptivity under wide angles of incident. At last, the absorptivity under different temperatures are investigated for evaluating the performance of the proposed absorber. All the good performance indicates the designed metasurface absorber promising in different applications, such as Radar Cross-Section reduction, low cost anechoic chambers and stealth technology.

## 2. Design of cylindrical-water-resonator-based absorber

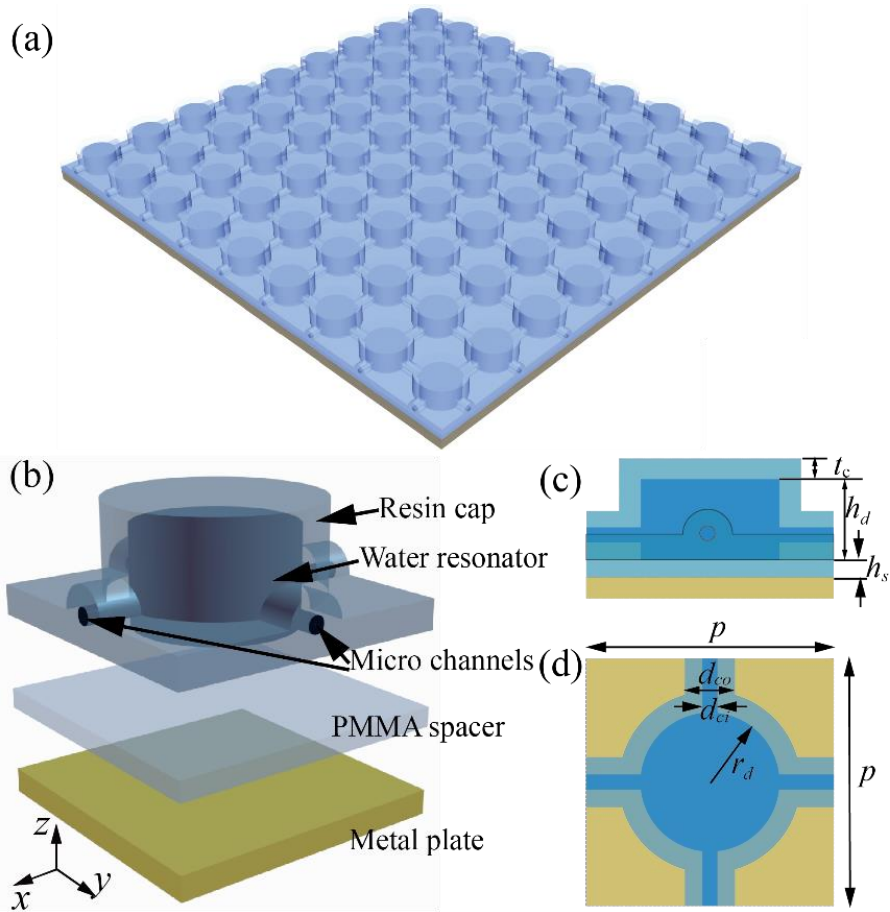

Figure 1 (a) Schematic diagram of the cylindrical-water-resonator-based ultra-broadband absorber. (b) Prospective view of unit cell of the absorber. Each part of the unit cell is ranged layer by layer. (c) Side view and (d) top view of the unit cell. The geometrical parameters are as follows: $p = 11.4$ mm, $r_d = 3.2$ mm, $h_d = 3.8$ mm, $t_c = 1$ mm, $h_s = 0.8$ mm, $d_{ci} = 0.6$ mm, $d_{co} = 2.2$ mm.

The illustration of the cylindrical-water-resonator-based absorber is shown in Figure 1a, and Figure 1b gives the perspective review of each single element. Figure 1c and 1d shows the side view and top view of the unit cell of the absorber, respectively, and dimensions of the unit cell are marked in the figures. The unit cell is periodically extended in the $x$ and $y$ directions with a lattice period of $p$. Each cell of the absorber consists of a water cylinder placed between a resin cylindrical cap and a bottom space layer made by Polymethyl methacrylate (PMMA). The water cylinder has a radius of $r_d$ and height of $h_d$ while the height of PMMA spacer is denoted as $h_s$. A metallic layer, copper here, is bonded to the PMMA spacer's backside, as the backplate of the whole element. The resin cylindrical cap and bottom spacer layer form the container of the water resonator. In order to inject water to the container, two microchannels are added in the resin cap, with a diameter of $d_{ci} = 0.7$mm. As the size of microchannels is much smaller than operating wavelength, their effect can be neglected. By optimizing the size of the water resonator and the height of PMMA space layer, good impedance matching between air and absorber can be obtained and near unit absorption appears over the broadband because of the high dielectric loss of water.

3. **Numerical analysis and measured results**

In order to investigate the working principle of the proposed absorber and optimize the design, numerical analysis is firstly carried out. The commercial full-wave simulation software CST Microwave Studio based on Finite Integration Technique (FIT) is used for numerical simulation. In the simulation, the unit cell boundary is used and the incident EM wave is assumed to propagate from the $+z$ axis. As the symmetry of the structure, the absorber is independent to the polarization of the incident wave and here $x$-polarized wave is assumed. For absorbers with metal backed, the absorptivity in simulation can be defined as $A(\omega) = 1 - R(\omega)$, where $R(\omega)$ represents reflectivity and can be calculated through the simulated $S$-parameters.

The dielectric of the resin and PMMA for container of the water resonators are set as $\varepsilon_r = 3 \times (1-j0.01)$ and $\varepsilon_r = 2.55 \times (1-j0.001)$, respectively. To represent the microwave dielectric characteristic of the pure water, Debye model is used in numerical simulation. Figure 2 shows the dielectric constant of pure water at ambient temperature. With reference to the figure, two things are worthy to be noticed. Firstly, the frequency-dependent characteristic of water's dielectric constant can be obviously observed. The real part of $\varepsilon$ shows a falling trend and it varies from about 74 to 29 during the band of interest. In the all-dielectric resonator design, the operating bandwidth of certain mode is usually narrow as the electrical size of the structure varies with frequency. However, when the permittivity of the material also changes with frequency, this issue may be overcome. On the ground of this, the changing of water's dielectric constant provides us an opportunity to design water-based all-dielectric resonator with broad mode bandwidth. Secondly, it can be seen that the imaginary part of dielectric constant keeps at a high level, indicating the loss of water is quite high over the whole band.

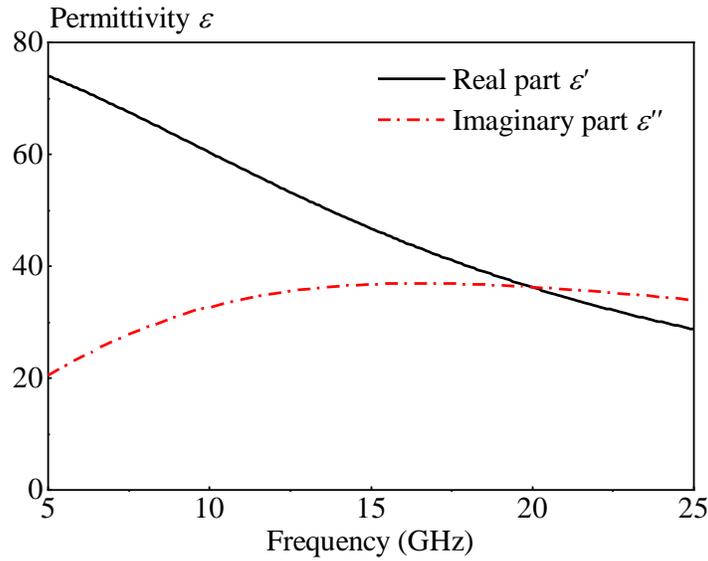

Figure 2 Permittivity of water during 5-25 GHz obtained using Debye model.

The simulated absorption spectrum of the proposed cylindrical-water-resonator-based absorber is shown in Fig. 3(a) with black line. The final optimized dimensions are also given in Fig. 1. High absorptivity can be observed over a wide operating band. The absorptivity of the proposed absorber is higher than 90% over the whole operating band from 5.58 to 24.21 GHz and the relative bandwidth can reach as high as 125%. In the operating band, four peak values can be found at 6.07, 8.10, 16.19 and 23.45GHz, which has been marked as $f_1$ to $f_4$. The absorptivity at these frequencies are 99.15%, 96.95%, 98.89% and 94.77%, respectively. The four absorption peaks indicate four resonances at these frequencies, in which the lower frequency points $f_1$ and $f_2$ correspond to the dielectric resonator (DR) mode while the higher frequency points $f_3$ and $f_4$ correspond to the spoof SPP mode, which will be analyzed in detailed later. As the loss of water is quite high over the band of interest, another two water-based absorbers are also simulated for comparison. Case I is a water plate with the same height as that in the proposed absorber and a metal plate is also placed on back of the water plate. Case II is the water resonator without metal backed. In this case, the size of the water resonator and the container is also the same with that in the proposed absorber. With reference to Figure 3(a), we can see that for Case I, water plate can also absorb the incident wave, but the absorptivity is low, varying only between 0.3-0.45. This is mainly caused by the mismatch between water plate and air as the dielectric of water is quite high compared with air. Water resonator without metal backed used in Case II shows different characteristic. An upward trend with frequency increases can be observed and the absorptivity varies from about 0.2 to 0.87. Two local peaks appear at 6.33 GHz and 20.88 GHz, corresponding to the DR mode and grating mode of the water resonator, respectively. As there is no metal backed, the SPP mode cannot be excited for Case II. However, the absorptivity is relative low over the whole band and the highest value is only 0.869, which is not enough for microwave absorption. This comparison can prove that the high absorption is mainly caused by the strong resonance of the cylindrical water resonator.

To fully understand the operating principle of the proposed absorber, the simulated normalized input impedance calculated by *S*-parameters is plotted in Fig. 3(b). From the figure, one can see that at the four resonant points, $f_1$ to $f_4$, the imaginary part Z′ is near zero and the real part is near unit, indicating the absorber owns good impedance matching with air and the wave reflection at the absorber-air interface is quite low. The good impedance matching ensures that the energy enters the absorber and then being totally absorbed owning to the high dielectric loss of water.

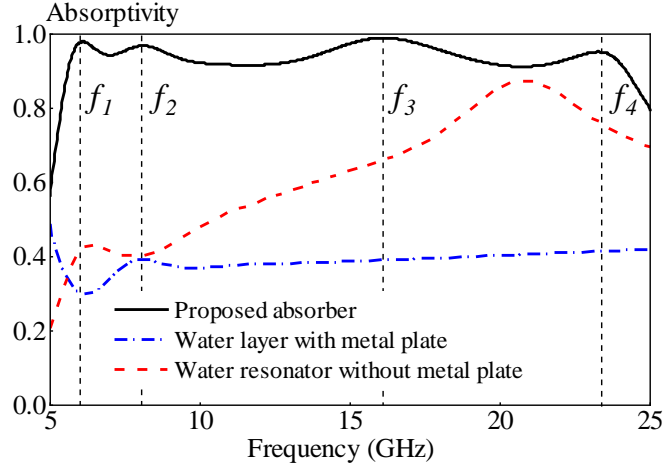

(a)

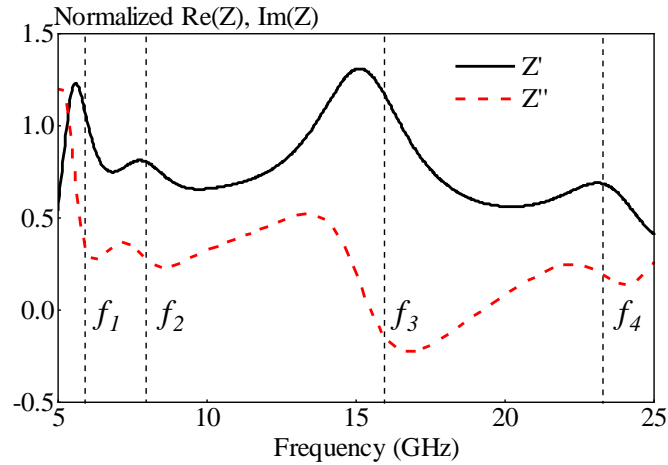

(b)

Figure 3 (a) Simulated absorption spectra of proposed water-resonator-based absorber, water plate with metal backed and cylindrical water resonator without metal backed. Four absorption peaks of the proposed absorber have been marked as $f_1$, $f_2$, $f_3$, and $f_4$ in figure. (b)Simulated input impedance of water-resonator-based absorber. $f_1$, $f_2$, $f_3$, and $f_4$ are also marked in figure.

The vector electric field (E-field) and magnetic field (H-field) in the resonator are also simulated to further investigate the working principle and the results are given in Figure 4. In the simulation, the polarization of incident wave is set as *x*-polarization. From the results in Figures 4(a) and (b), a looped E-field distribution can be obviously found in the water resonator at $f_1 = 6.07$ GHz and the *H*-field shows up a dipole type, verifying the water resonator is excited in its DR mode, magnetic dipole mode. Comparing the field distribution at $f_2 = 8.10$ GHz with that at $f_1$, quite similar manner can be observed, indicating the similar resonant modes. This can be explained using the dielectric constant of water shown in Figure 2. For a dielectric resonator with fixed dimensions, the resonating frequency for a certain mode in the resonator can be considered owning a linear relation with its electrical size, which can be determined using $\lambda/\sqrt{\varepsilon}$, where $\lambda$ is the wavelength in free space. With reference to Figure 2, it can be seen that $\varepsilon$ decreases with frequency increasing, resulting in the electrical size of the resonator varied very small over a wide band. This feature allows the DR mode maintains over a wide frequency band, manifested by multiple resonance points in the band, namely $f_1$ and $f_2$. In addition, it also can be observed that at $f_1$ and $f_2$, the incident wave is strongly confined in the water resonator and the high absorption is thus caused by the high dielectric loss of water. Thus the changing of water's dielectric constant can eliminate the effect caused by the changing of electrical size of the resonator with frequency. At lower frequency $f_1$, the water owns higher dielectric constant while at $f_2$, the dielectric constant decreases significantly.

Next, the working principle in higher frequency, $f_3$ and $f_4$, is analyzed. As is shown in Figures 4(e) to (h), it can be seen that the incident wave is mainly confined between the water resonator and metal plate. In this case, the array of the water resonators works as a grating and the spoof SPP mode that usually existed between dielectric and metal can be excited. This is similar to the excited mode in [45] and [48], where the spoof SPP mode is excited using hemispherical water resonator. As the frequency-dependent dielectric constant of water, the operating bandwidth of dielectric grating is wide enough that spoof SPP mode can be excited at different frequency points. By combining the multiband of dielectric mode and spoof SPP mode, an ultra-broadband with high absorptivity can be obtained using the proposed absorber.

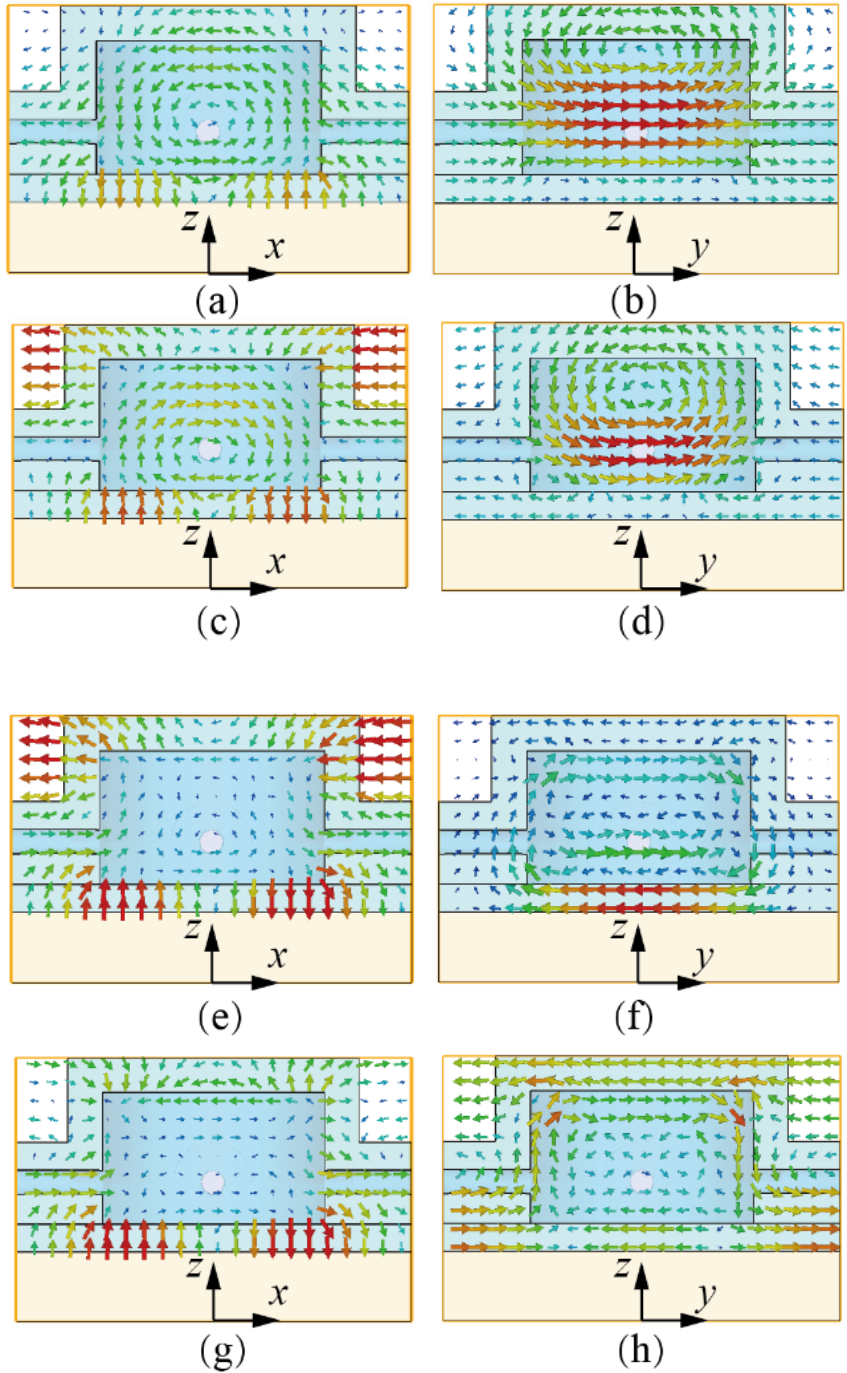

Figure 4 Simulated vector field distribution in the absorber. (a) *E*-field in *xoz* plane for $f_1 = 6.07$ GHz. (b) *H*-field in *yoz* plane for $f_1 = 6.07$ GHz. (c) *E*-field in *xoz* plane for $f_2 = 8.1$ GHz. (d) *H*-field in *yoz* plane for $f_2 = 8.1$ GHz. (e) *E*-field in *xoz* plane for $f_3 = 16.19$ GHz. (f) *E*-field in *yoz* plane for $f_3 = 16.19$ GHz. (g) *E*-field in *xoz* plane for $f_4 = 23.45$ GHz. (h) *H*-field in *yoz* plane for $f_4 = 23.45$ GHz.

To verify the design, the prototype of the proposed absorber containing 15×15 unit cells shown in Fig. 1 is fabricated and measured. The total size of the absorber is 181×181 mm². 3D printing technology is used to manufacture the top resin cap and the PMMA are bounded with the resin cap, forming the whole container for the water resonator. A copper foil is stucked on the back side of the PMMA, acting as the metal plate. The water is injected through the microchannel. The inset in Figure 5 shows the prototype of the absorber. In order to investigate the absorption performance of the proposed absorber, the reflection coefficient of the absorber is measured using vector network analyzer with two broadband horn antennas. A metal plate with the same size is also measured using the same method as reference and the reflectivity of the absorber is normalized to that of metal plate. The simulated and measured results are given in Figure 5. Good agreement between simulation and measurement can be observed from the results. In most of the band of interest, the absorptivity is higher than 90%. Some slight difference may be caused by the imperfections in fabrication process and the measurement error.

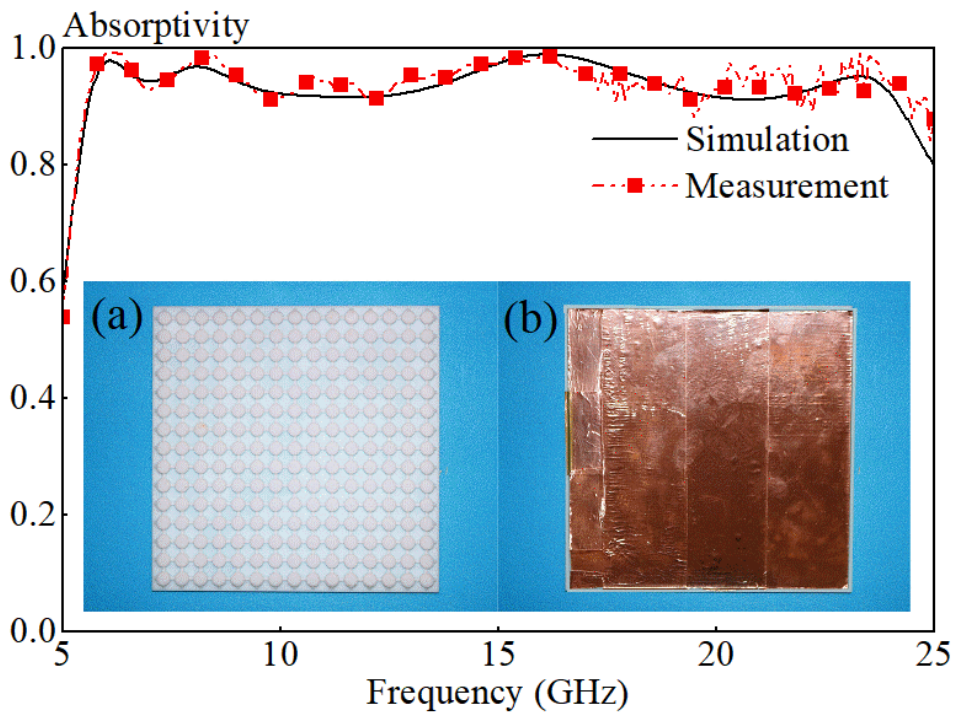

Figure 5 Simulated and measured absorption spectra of the proposed absorber. The inset: photo of prototype of the absorber with 15×15 elements. (a) Top view. (b) Bottom view.

For fully evaluating the performance of the cylindrical-resonator-water-based absorber, the absorptivity under EM waves with different polarizations and incident angles are simulated and the results are shown in Figure 6. TE and TM incidents are both considered and the incident angle varies from 0° to 45°. It can be seen from the figure that for TM incident, the absorptivity keeps higher than 90% over the whole operating frequency when incident angles below 30°. When the incident angle increases to 45°, the absorptivity is higher than 90% in most of the frequency band and higher than 85.5% over the rest band. For the case of TE incident, as incident angle increases, the absorptivity decreases at lower frequency band while keeps higher than 90% at higher frequency band. For the incident angle below 45°, the absorptivity is higher than 77.6% over the whole operating frequency band.

It is worth noting that there is significant difference between TE and TM incident in Figure 6, which is consistent with some previous work, such as Ref. [44]. This phenomenon can be understood with the help of working principle. As mentioned above, at lower frequency band, DR mode, i.e. magnetic dipole mode, of the water resonator is excited and this mode is quite sensitive to the change of the intensity of incident magnetic field. For TE mode incident, when the incident angle increases, the magnetic field will decrease and this will make the DR mode weaker, thus the absorptivity decreases significantly. However, the spoof SPP modes at higher frequency band are not so sensitive to magnetic field and as the incident angle increases, the absorptivity will keep at a high level. This is consistent with the results shown in Figure 6(a). For TM mode incident, the change of magnetic field is not small and the electric field will change significantly with increased incident angle. Since the DR mode and the spoof SPP mode are not sensitive to electric field, the absorptivity remains quite high. Based on the simulated results under different incident angles, the absorber shows good performance over a wide incident angle and broad operating band.

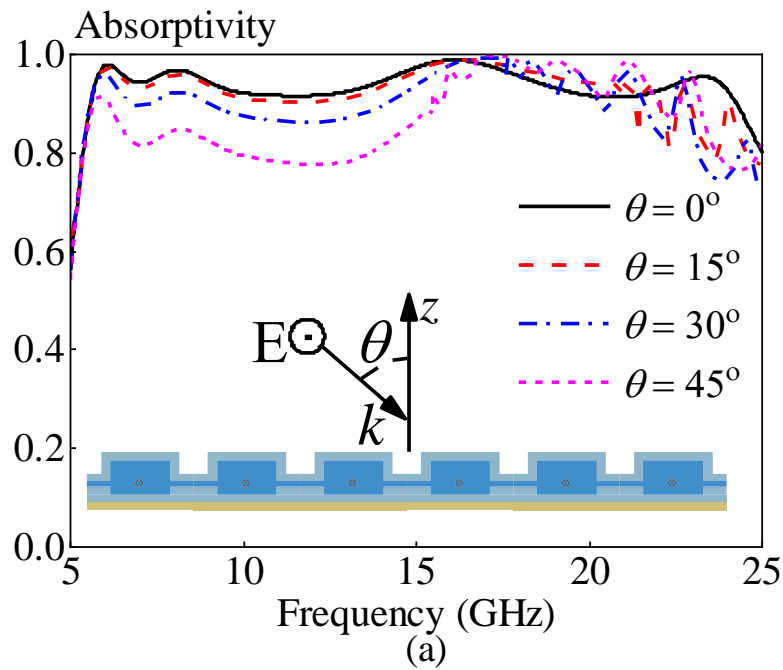

(a)

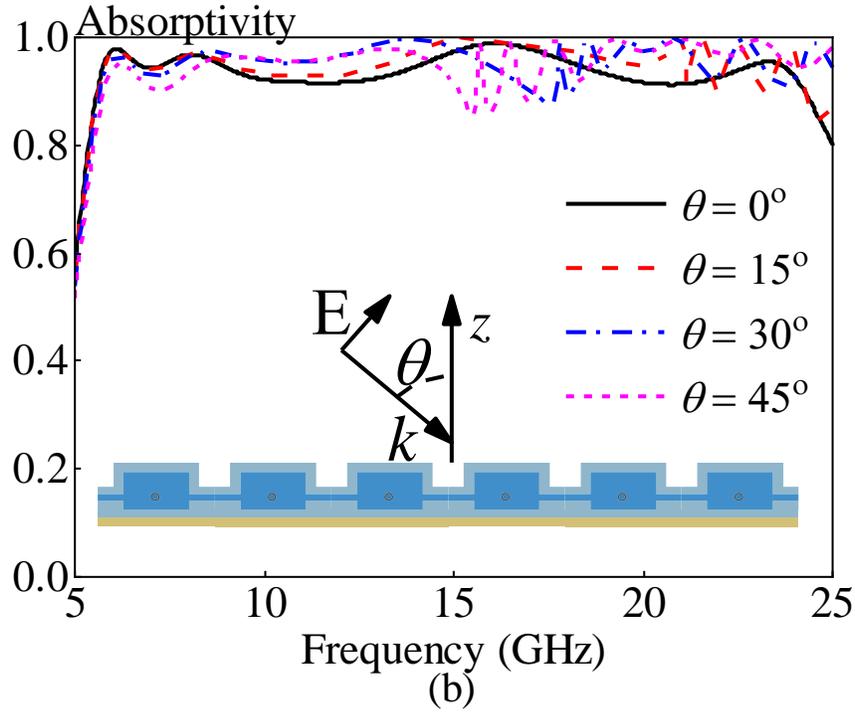

Figure 6 Absorption spectra of the proposed water absorber for oblique incidence waves with different incident angles and different polarization states. (a) TM mode and (b) TE mode.

As an absorber for transferring EM energy to thermal energy, the temperature of the water may be changed during the process of application. Meanwhile, water is a temperature sensitive material and the dielectric constant changes with temperature. So, it is worthy to investigate the absorptivity under different temperatures and discuss the temperature stability of the absorber. The dielectric constant of the water under different temperature can be predicted using the Debye formula [34, 44] and the effect of the resin container and PMMA plate is not considered in the simulation. Figure 7 shows the absorption spectrum with temperature varies from 0° to 100°. With reference to the figure, one can see that the change of the absorptivity under different temperature condition is quite small over the whole band of interest. When the temperature varies between 0° and 40°, the absorptivity keeps higher than 90% from 5.58 to 24.21 GHz. When the temperature increases to 60°, the absorptivity decreases slightly at lower frequency band, but still keeps higher than 87% over the operating band. When the temperature continues to increase, the absorptivity is higher than 78% over most part of the operating band, except for some frequency point at lower frequency band. There is reason to believe that the proposed absorber is weakly depended on the water temperature and it can work steadily when the temperature changes significantly.

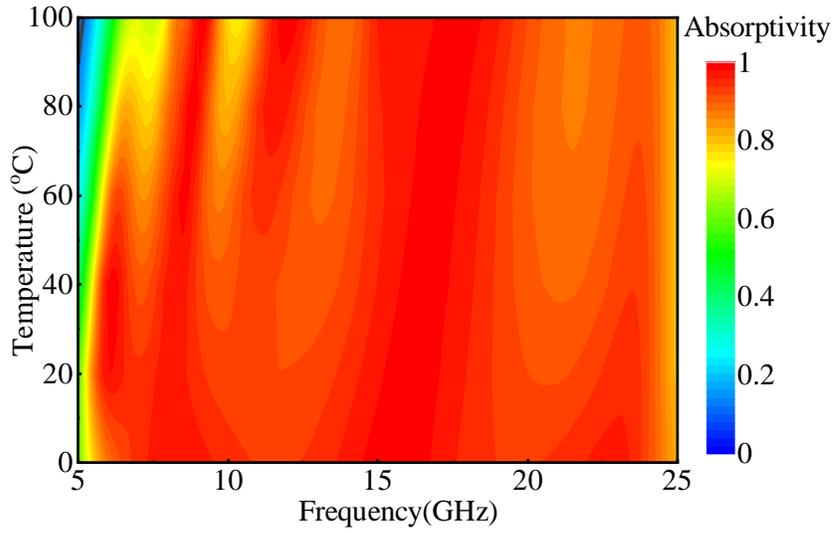

Figure 7 Absorption spectrum of the absorber at different temperatures.

## 4. Discussion

Water has been widely used in design of low cost absorber with high absorption for its high dielectric loss recently. This includes using the water plate as dielectric substrate and the water resonator with different shapes, such as droplet and cub. To highlight the merits of the absorber presented here, a comparison with the previous water-based absorbers is made and the results are summarized in Table I. From the table, it is clearly that the absorption bandwidth of the proposed absorber is the widest, with the value as high as 125%.

| Ref. | Absorber Type | Operating Band | Relative Bandwidth |
| --- | --- | --- | --- |
| [38] | Water resonator with droplet shape | 8-18 GHz | 76.9% |
| [40] | Water plate with grid shape | 8.1-22.9 GHz | 95.5% |
| [41] | Water as substrate | 6.2-19 GHz | 101.6% |
| [42] | Water resonator with droplet shape | ≈17.2-40 GHz | 78.9% |
| [43] | Water resonator with rectangular shape | 8.96-12.0 GHz | 28.8% |
| [44] | Water plate with hole array | 12-29.6 GHz | 84.6% |
| [45] | Water resonator with droplet shape | 7.5-15GHz | 66.7% |
| [45] | Water cube | 4.5-8 GHz | 56% |
| This work | Water resonator with cylindrical shape | 5.58-24.1GHz | 125% |

## 5. Conclusion

In summary, a cylindrical-water-resonator-based absorber with an ultra-broad operating band is proposed in this paper. At lower frequency band, the DR mode of the water resonator is excited while the spoof SPP mode based on grating effect is excited at higher frequency band. Benefit from the advantage of the dielectric constant of water varying with frequency, the operating bandwidth of these two modes is quite wide. By optimizing the design, ultra-broad operating band can be obtained with near-unit absorptivity. The measured results show that the absorptivity of the designed absorber is as high as 90% over almost the whole operating band from 5.55 GHz to 24.1 GHz and the relative bandwidth reaches 125%. The absorptivity under different polarized incident waves and different incident angles are simulated for in-depth exploration. In addition, the thermal stability is also investigated for evaluating the performance of the absorber. All the good performance of the absorber makes it promising in different applications, such as the low cost anechoic chambers and stealth technology.